\documentclass[sigconf]{acmart}
\usepackage{multirow,subcaption,makecell}

\AtBeginDocument{%
  \providecommand\BibTeX{{%
    \normalfont B\kern-0.5em{\scshape i\kern-0.25em b}\kern-0.8em\TeX}}}

\copyrightyear{2023} 
\acmYear{2023} 
\setcopyright{acmlicensed}\acmConference[SIGIR-AP '23]{Annual International ACM SIGIR Conference on Research and Development in Information Retrieval in the Asia Pacific Region}{November 26--28, 2023}{Beijing, China}
\acmBooktitle{Annual International ACM SIGIR Conference on Research and Development in Information Retrieval in the Asia Pacific Region (SIGIR-AP '23), November 26--28, 2023, Beijing, China}
\acmPrice{15.00}
\acmDOI{10.1145/3624918.3625321}
\acmISBN{979-8-4007-0408-6/23/11}

\begin{document}

\title{Investigating the Influence of Legal Case Retrieval Systems on Users' Decision Process}



\author{Beining Wang}
\affiliation{%
  \institution{Quan Cheng Laboratory, \\
Dept. of CS\&T, Institute for Internet Judiciary, Tsinghua University}
  \city{Beijing}
  \country{China}
  \postcode{100084}
}
\email{Benson0704@outlook.com}

\author{Ruizhe Zhang}
\affiliation{%
  \institution{Quan Cheng Laboratory, \\
Dept. of CS\&T, Institute for Internet Judiciary, Tsinghua University}
  \city{Beijing}
  \country{China}
  \postcode{100084}
}
\email{u@thusaac.com}

\author{Yueyue Wu}
\affiliation{%
  \institution{Quan Cheng Laboratory, \\
Dept. of CS\&T, Institute for Internet Judiciary, Tsinghua University}
  \city{Beijing}
  \country{China}
  \postcode{100084}
}
\email{wuyueyue@mail.tsinghua.edu.cn}

\author{Qingyao Ai}\authornote{Corresponding author}
\affiliation{%
  \institution{Quan Cheng Laboratory, \\
Dept. of CS\&T, Institute for Internet Judiciary, Tsinghua University}
  \city{Beijing}
  \country{China}
  \postcode{100084}
}
\email{aiqy@tsinghua.edu.cn}

\author{Min Zhang}
\affiliation{%
  \institution{Quan Cheng Laboratory, \\
Dept. of CS\&T, Institute for Internet Judiciary, Tsinghua University}
  \city{Beijing}
  \country{China}
  \postcode{100084}
}
\email{z-m@tsinghua.edu.cn}

\author{Yiqun Liu}
\affiliation{%
  \institution{Quan Cheng Laboratory, \\
Dept. of CS\&T, Institute for Internet Judiciary, Tsinghua University}
  \city{Beijing}
  \country{China}
  \postcode{100084}
}
\email{yiqunliu@tsinghua.edu.cn}

\renewcommand{\shortauthors}{Wang, et al.}

\begin{abstract}
  Given a specific query case, legal case retrieval systems aim to retrieve a set of case documents relevant to the case at hand. Previous studies on user behavior analysis have shown that information retrieval (IR) systems can significantly influence users' decisions by presenting results in varying orders and formats. However, whether such influence exists in legal case retrieval remains largely unknown. This study presents the first investigation into the influence of legal case retrieval systems on the decision-making process of legal users. We conducted an online user study involving more than ninety participants, and our findings suggest that the result distribution of legal case retrieval systems indeed affects users' judgements on the sentences in cases. Notably, when users are presented with biased results that involve harsher sentences, they tend to impose harsher sentences on the current case as well. This research highlights the importance of optimizing the unbiasedness of legal case retrieval systems.
\end{abstract}

\ccsdesc[500]{Information systems~Information retrieval}
\ccsdesc[300]{Information systems~Presentation of retrieval results}



\keywords{Legal Case Retrieval, Judicial Judgement, User Study}


\maketitle

\section{INTRODUCTION}

Given a specific query case, legal case retrieval systems aim to retrieve a set of case documents relevant to the case at hand. By providing access to a vast collection of past judgments and their associated legal reasoning, they contribute to the development and application of consistent legal principles, ensuring fairness, predictability, and uniformity in legal decision-making. Therefore, legal case retrieval systems have long served as an essential part of both the common law and civil law systems. For example, in China (a country with civil law systems), the Supreme People’s Court explicitly requires all judges to submit relevant cases retrieved by legal case retrieval systems to the court before making the final decision on a specific case.

Previous studies on user behavior analysis have shown that information retrieval (IR) systems can significantly influence users' decisions by presenting results in varying orders and formats. Because search engines and recommendation systems have served as the major entries, if not the only entries, for people to access information on the internet, how these systems select and organize retrieved results could directly affect what and how information is distributed to downstream users.
User’s opinions and decisions are thus likely to be influenced by the IR system’s behaviors and strategies.
For example, previous studies have shown that people’s views on certain topics could be significantly different when we present different results on the top of search engine result pages (SERPs) ~\cite{AJIM2019}.

To this end, we present one of the first studies on how user decisions are influenced by the retrieved results of legal case retrieval systems.
Our goal is to provide a quantitative analysis on how the distribution of sentences in the retrieved cases would affect users' decisions on the sentences of the query cases.
Specifically, we conduct experiments on a Chinese legal case retrieval dataset, i.e., LeCaRDv2, and recruited 94 participants who major in criminal law (82 of which have already passed National Judicial Exam of China) to simulate judges in courts who need to make decisions on query cases based on the fact descriptions and cases retrieved by legal case retrieval systems.
We carefully crafted the retrieved case lists for each participant to study whether the changes of retrieval results would affect their final decisions on the query cases.
Experimental results show that participants tend to impose more penalties to the accused when they are presented with cases with harsher sentences, despite of the relevance of the presented cases.
This highlight the importance of unbiasedness in the optimization of legal case retrieval systems.

Since the fact that legal case retrieval systems have become increasingly prevalent in the past years, studying whether legal case retrieval systems have an influence on users in the field of justice and evaluating this influence can help eliminate biases in legal case retrieval systems and optimize fairness and justice in the field of justice, with far-reaching practical significance and application value.

The remainder of the paper is organized as follows. Section 2 reviews the related works on the influence of information retrieval systems, the bias in information retrieval systems and legal case retrieval systems. Section 3 presents details of the user study we conduct. Section 4 describes the results of the user study. The last section concludes the paper.
\section{RELATED WORK}
Here are some studies on related topics such as the influence of information retrieval systems on their users, the bias in these systems, and the current state of legal case retrieval systems.
\subsection{IR system's influence on user behaviors.}
Information retrieval systems, including search engine systems and recommendation systems, have a certain influence on users. For example, previous studies show the the use of search engine may make people over-confident~\cite{PNAS2021mistake} and recommendation systems in online market may induce user to purchase more often ~\cite{JMIS2010sale,ISR2020impact}. Studies on this aspect are abundant. 

Some companies have borrowed the idea that recommendation systems have an influence on users' decision-making process and achieved certain results, for example, previous study shows that simply changing the video ranking to place it before popular videos can make it more popular~\cite{IMC2010rank}. Search engine systems also have an influence on users' decision-making process and cognition. For example, it is found that when the information retrieved by search engine is similar to users' cognition, they are likely to make the same decision ~\cite{PEC2014,UMAP2021} .In the job market, a previous study shows that relevant searches for jobs affect users' decision-making when looking for jobs~\cite{WSDM2020job}. In the study of search engine result pages, it is pointed out that just viewing the top entries returned by the search engine can change users' views on the search topic~\cite{socialmedia}. Similarly, it is shown that it is feasible to manipulate search results to change voters' preferences in a undecided democratic election~\cite{PNAS2015}. Furthermore, there is a study show that even a brief selective exposure to online search results can affect users' attitudes towards elections~\cite{JCMC2015}.
\subsection{The Bias in IR Systems}
There is various bias types in information retrieval systems ~\cite{CHIIR2021} and there are previous studies focus on the bias itself and its influence attempt to address it and reduce it ~\cite{CSCW2017,SIGMODrecord2017,ICTIR2015,SIGIR2020}. A previous study also points out biases present in search engine result pages~\cite{CHIIR2017bias}. Additionally, previous studies in the medical diagnosis field show that this bias has both positive and negative effects~\cite{ICTIR2017search}. On the user side, biases may arise due to the biases of information retrieval systems. There is a study find cognitive biases in users when using information retrieval systems by asking them to answer questions before and after searching for information and labeling the usefulness of the searched information~\cite{JAMIA2007}. Previous studies find that users are very uncertain when evaluating the credibility of search engine results pages and often have different impressions from fact-checking~\cite{AJIM2019}. Previous work demonstrated that search engine result pages can cause users to have biases~\cite{CHIIR2017bias}, and it is pointed out that this bias can also affect users' information retrieval behavior in return~\cite{JISTAP2020}. If considering manipulating information retrieval systems, according to the research, although such behavior can influence users' behavior, users can detect that the search results have been manipulated~\cite{CHI2003}.
\subsection{Legal Case Retrieval Systems}

The work published in the top-level comparative law journal in the United States\textit{American Journal of Comparative Law}, indicates that although China, as a non-precedent jurisdiction, explicitly prohibits judges from referencing prior cases as the basis for their judgements, and in previous investigations, most judges have made clear that prior cases play only a marginal role in their decisions, some Chinese judges still rely heavily on cases retrieved when making decisions~\cite{2021Precedents}. Since the fact that some judges in China, a statutory law country, rely on the cases retrieved, a credible and well-performing case retrieval system is essential.

There are some previous studies on legal case retrieval systems. The topics are various. One of these studies constructs a large legal case retrieval dataset ~\cite{lecard1}. While previous studies investigate the user behavior in legal case retrieval to benefit the design of the corresponding retrieval systems to support legal practitioners ~\cite{Shao2021InvestigatingUB}, there are also some studies focus on the ranking to improve the performance of legal case retrieval systems ~\cite{2020BERT,2022masigir}.

\section{USER STUDY}
Initially, we clarify our research question. After that, we introduce our experiment platform and online user study, including how we recruit participants, design task data, and collect experiment results.

\subsection{Research Questions}
The research questions we want to address in this paper are:

\begin{itemize}
    \item Would user's judgments be affected by the distribution of cases retrieved by the legal case retrieval systems?
    \item If so, are there any patterns between the distributions of the retrieval results and the user's sentences to the query cases?
\end{itemize}

These two questions are the key to understand whether legal case retrieval systems have an influence on their users. If there are special patterns between case retrieval results and user's judgments on query cases, we could use them to identify the potential bias in existing retrieval systems and improve the overall quality and fairness of the legal judgment process.

\subsection{Methods}
\begin{figure}[t]
  \centering
  \includegraphics[width=\linewidth]{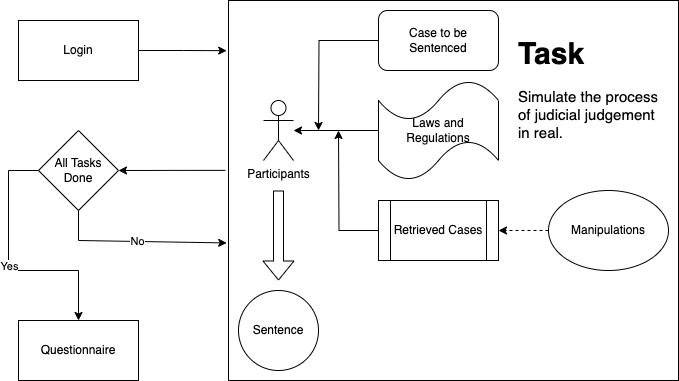}
  
  \caption{The flowchart of user study}
  \label{fig:flow}
  \Description{Note that the perturbation does not exist in real.}
\end{figure}

The workflow of our experiment is shown in \textbf{Figure~\ref{fig:flow}}. 
The experiment consists of four tasks and one user questionnaire, with each task requiring participants provide sentences to the query case based on the information (i.e., law descriptions and cases retrieved by the retrieval systems) provided by the platform and their own thinking and cognition. The questions in the questionnaire are listed in \textbf{Table~\ref{tab:questionnaire}}, and the possible answers to each question is \textit{completely disagree}, \textit{disagree}, \textit{neutral}, \textit{agree}, and \textit{completely agree}. After completing all tasks, participants are required to complete this questionnaire based on their experiment experience. The design of the experiment tasks simulates the process of judicial judgement in a real judicial judgement scenario. In practice, after conviction, a judge would make a sentence based on the following information:
\begin{itemize}
    \item Related laws and regulations, including criminal law and sentencing guidelines.
    \item "Extraneous factors" beyond the case circumstances, including the attitude of the suspect towards confession, the economic situation of the victim, public opinion, and social stability factors.
    \item Sentences of relevant cases retrieved by the legal case retrieval systems.
\end{itemize}

\begin{table*}
  \caption{Questionnaire content}
  \label{tab:questionnaire}
  \begin{tabular}{ll}
    \toprule
   Question Topic&Question Content$\bullet$\\
    \midrule
     Satisfaction&You feel satisfied with the overall experiment.\\
    Confidence& You have confidence in your sentence.\\
    Effort& \makecell[l]{The completion of the task requires a significant amount of mental effort\\ (e.g., calculation, thinking, decision-making, and memory) ~\cite{HART1988139}.}\\
    Overall Unbiasedness& The cases retrieved provided for the task are fair.\\
    Gender Unbiasedness& \makecell[l]{When providing the cases retrieved, the system tends to assign harsher sentences to \\male defendants (or female defendants).}\\
    Age Unbiasedness& \makecell[l]{When providing the cases retrieved, the system tends to assign harsher sentences to \\younger defendants (or older defendants).}\\
    Region Unbiasedness& \makecell[l]{When providing the cases retrieved, the system tends to assign harsher sentences to\\ defendants from certain regions.}\\
    Usefulness& The cases retrieved assist you in completing the task.\\
  \bottomrule
\multicolumn{2}{l}{\small $\bullet$ Answers are chosen from completely disagree, disagree, neutral, agree and completely agree.}\\
\end{tabular}
\end{table*}
Note that a standard judicial judgment process usually consists of two separate parts, i.e., conviction and sentence. For simplicity, we only focus on the sentence part and reveal the actual crime of the query case and corresponding codes to the participants directly in the beginning of each task. Therefore, the participants only need to focus on providing a sentence to the case based on the information at hand.

In the experiment, participants are divided into different groups. Specifically, all participants need to provide sentences to the same four query cases used in the four experiment tasks, but we manipulated the distribution of the sentences in the retrieved cases shown to different groups of participants. Through this way, we aim to examine whether the distribution of the retrieved cases would affect participants opinions on each query cases.

\subsection{Dataset}
The dataset used in this experiment is LeCaRDv2\cite{lecard2}. The dataset includes several criminal verdicts in China and is organized by query. For each query, the top 30 documents of BM25 is retrieved from a corpus with over 55,000 documents. Overall, the dataset has 800 queriess and the average BM25 total score of these top 30 documents is 20.89. 

For simplicity, we picked 4 representative queries from the 800 queries for our experiments, and the process of query selection is described as the followings. To begin with, we choose specific 4 crimes, that is, injury, robbery, rape and murder, which are 4 violent serious crimes. For these crimes, at the sentencing level, the range of sentence that are applicable under the criminal code is broad, which is suitable for observing the influence of a case retrieval system on judicial judgement made by users. Additionally, we select only one query for each of these crimes, in order to prevent participants from being influenced by each other if they sentence two different cases with the same crime later on, thereby making it inconvenient to group tasks at the level of trials.

To pick these 4 queries out, we filter the queries that
\begin{enumerate}
    \item are not in those crimes
    \item are longer than 5,000 words
    \item are sentenced to probation or death
    \item their documents candidates that with the same crime and not sentenced to probation or death are more than 10
\end{enumerate} 
Then, for each crime, if its queries are more than one, we choose the median sentence query. The overview of the picked 4 queries are showned in \textbf{Table~\ref{tab:task queries}}
.
\begin{table}
  \caption{Overview of task cases}
  \label{tab:task queries}
  \begin{tabular}{ccl}
    \toprule
    Crime&Original Sentence (months)& Document Length\\
    \midrule
    Injury & 7 & 1063 words\\
    Robbery & 42& 1160 words\\
    Rape & 39& 1333 words\\
    Murder & 156& 4951 words\\
  \bottomrule
\end{tabular}
\end{table}
\subsection{Participants}
Due to the high requirements of domain expertise in judicial judgment process, we only recruited participants with strong background in law. Specifically, we sent out invitations to law schools or firms and required participants to have a master's degree in criminal law or above, including master's degree students in criminal law currently studying, master's degree students expected to enter in the autumn of 2023, or doctoral degree students in criminal law. In total, we successfully recruited 94 law students as participants, more than 88\% of whom have passed National Judicial Examination in China.

\begin{figure}
    \centering
    \includegraphics[width=0.8\linewidth]{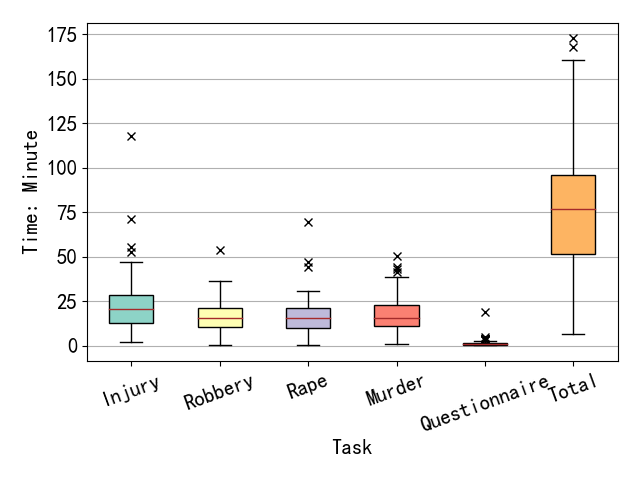}
    \caption{Experiment finish time distribution}
    \label{fig:time}
\end{figure}

Behavior logs of the participants collect such as clicking buttons or taking breaks. We conducted a pilot study and found it impossible for a participant to finish the whole experiment (including four tasks and a questionnaire) within 30 minutes if they take it seriously. As shown in \textbf{Figure~\ref{fig:time}}, the averaged time to finish the experiment is about 75 minutes. Therefore, we filtered out data from 5 participants with finish time less than 30 minutes.

For each task (query case), we provide three types of retrieval results: the \textbf{light group}, \textbf{control group}, the \textbf{harsh group}. Specifically, we first assume that the top 30 cases for each query case provided in the LeCaRDv2 dataset are relevant cases, which is reasonable because LeCaRDv2 has retrieved and ranked cases using a combination of multiple SOTA legal retrieval models. Then, we sort these cases based on the severity of the penalty in their sentences in increasing order. We further split the list into three set according to the order: the top 10 is the light set (which contains low-penalty sentence cases), the median 10 is the control set, and the last 10 is the harsh set (which contains heavy-penalty sentence cases). We randomly select cases from each set to form the light, control, and harsh group as described in \textbf{Table~\ref{tab:group}}. The key idea is to create three types of case lists for each query case that have different severity in terms of sentence penalty. Given the same query case, we can show different groups of cases to each participant to see whether the change of retrieved cases would affect people's judgments on the query case.

To prevent any potential coeffect between the manipulation of retrieval results in different tasks/query cases for the same participant, we randomly select groups for each task of each participant.

\begin{table}
  \caption{Overview of different groups}
  \label{tab:group}
  \begin{tabular}{lccc}
    \toprule
    Group&Light set&Median 
    set&Harsh set\\
    \midrule
    
    Light & 3&3&0\\
    Control  & 2&2&2\\
    Harsh  & 0&3&3\\
    
  \bottomrule
\end{tabular}
\end{table}

\section{RESULTS}

In order to address the research questions, we conduct a user study that aimed to examine the distribution of punishment across different groups. If significant differences are observed in the distribution of punishments across these groups, we would posit that the cases retrieved have an influence on users' sentence. To analyze the data, we first employ Levene's Test to assess the homogeneity of variance and subsequently use a two-sample $t$ test to determine significance across different groups.

To further bolster our findings, we utilize the results of the questionnaire to emphasize the validity of our experiment. If a majority of participants expresses high confidence in their sentences, it suggests that they are not aware of any manipulation and thus strengthens the credibility of our study. Similarly, if a majority reports that the cases retrieved are highly beneficial, this implies that their decision-making processes are influenced by the cases retrieved rather than solely relying on their knowledge and cognitive abilities, which further enhances the reliability of our experiment.

\subsection{User Study Results}
\begin{figure*}
    \centering
    \subcaptionbox{Injury}{\includegraphics[width=0.24\linewidth]{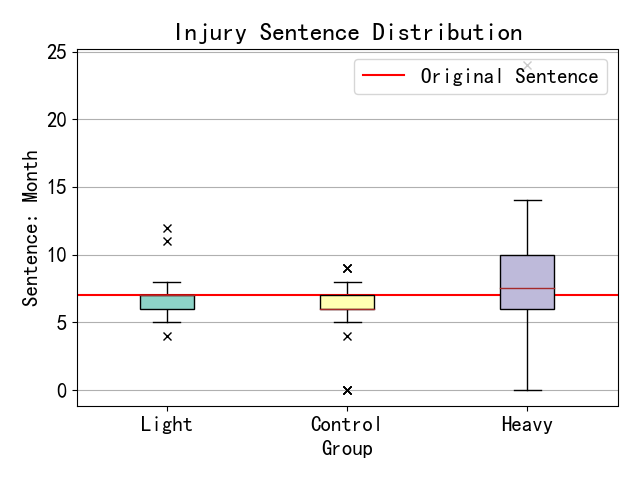}}
    \subcaptionbox{Robbery}{\includegraphics[width=0.24\linewidth]{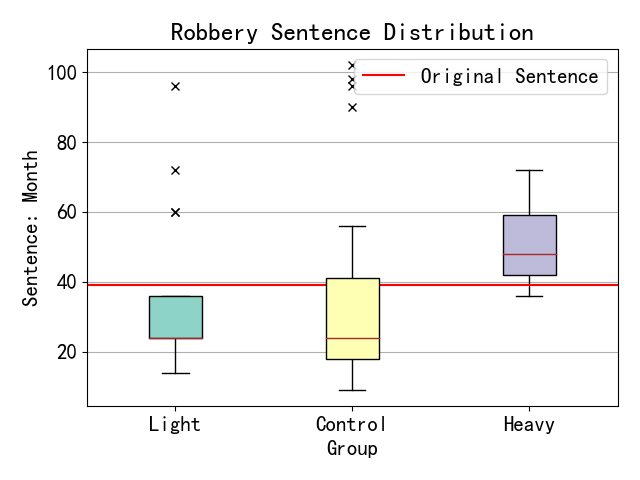}}
    \subcaptionbox{Rape}{\includegraphics[width=0.24\linewidth]{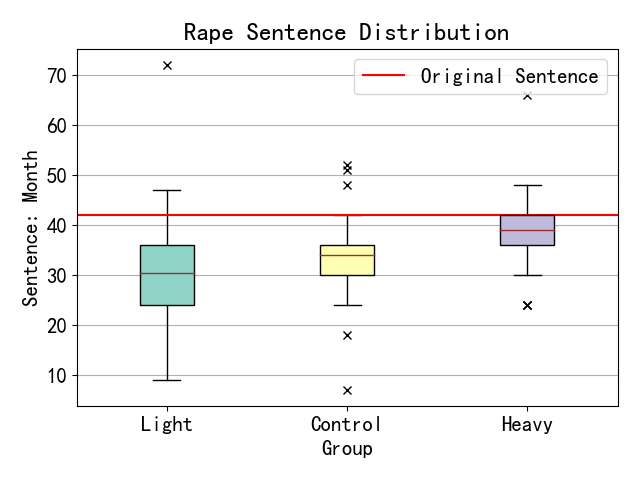}}
    \subcaptionbox{Murder}{\includegraphics[width=0.24\linewidth]{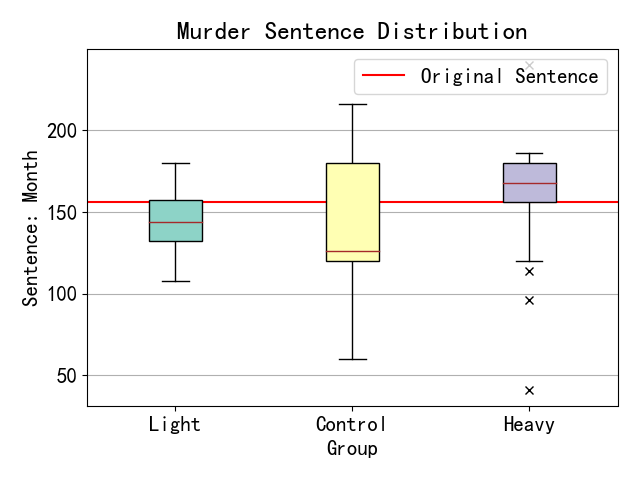}}
     \caption{Punishment distribution of different groups}
    \label{fig:distribution}
\end{figure*}

The sentence distribution of each task across different groups is shown in \textbf{Figure~\ref{fig:distribution}} and the sentences' statistics and significance are listed in \textbf{Table~\ref{tab:result}}.

Due to the fact that whether to reduce or not to reduce a punishment is within the discretion of judges in legal practice, in this experiment scenario, participants have made deviations within the scope of discretion due to the influence of cases retrieved. Moreover, the trend of deviation is consistent with that of participant grouping, indicating that the case retrieval system has influenced the decision-making process of participants in judicial judgement. Specifically, if participants are given reference to harsher punishments, they tend to make harsher judicial judgements, which is significant. However, if participants are given reference to lighter punishments, their tendency is not significant.

\begin{table*}
  \caption{Sentence results of different groups}
  \label{tab:result}
  \begin{tabular}{clcccl}
    \toprule
    Crime &Group & Participants & Median Sentence (months) & Average Sentence (months) & Significance ($P$ value)\\
    \midrule
    \multirow{3}{*}{Injury}&Light&27&7&6.74&0.0646 vs.  Control\\
    ~&Control&30&6&5.73&0.1293 vs.  Harsh\\
    ~&Harsh&32&7&7.28&0.5567 vs.  Light\\
    \hline
    \multirow{3}{*}{Robbery}&Light&30&24&32.93&0.4832 vs.  Control\\
    ~&Control&29&24&37.10&\textbf{0.0241} vs.  Harsh\\
    ~&Harsh&30&48&49.60&\textbf{0.0000} vs.  Light\\
    \hline
    \multirow{3}{*}{Rape}&Light&30&31&30.53&0.2052 vs.  Control\\
    ~&Control&31&34&34.00&\textbf{0.0422} vs.  Harsh\\
    ~&Harsh&28&39&38.68&\textbf{0.0048} vs.  Light\\
    \hline
    \multirow{3}{*}{Murder}&Light&30&144&136.53&0.1997 vs.  Control\\
    ~&Control&26&126&135.08&\textbf{0.0081} vs.  Harsh\\
    ~&Harsh&33&168&161.45&\textbf{0.0440} vs.  Light\\
    
    \bottomrule
  \end{tabular}
\end{table*}

In the task of intentional injury, it is observed that some participants in the experiment assigned the punishment of supervision or detention to the case details in the query. The number of cases with the punishments of supervision or detention is shown in \textbf{Table~\ref{tab:injury}}. It can be found that in the light punishment group, no participant made the choice of judicial judgement with the punishment of supervision or detention. However, in the control group and harsh punishment group, there are participants who assigned the punishment of supervision or detention, not just an isolated case. The number of participants with such assignments is significant. Furthermore, participants in the harsh punishment group tend to assign the punishment of supervision or detention to judicial judgements. This is a phenomenon worth further study.

\begin{table}
  \caption{Supervision or detention sentences in injury}
  \label{tab:injury}
  \begin{tabular}{lcc}
    \toprule
    Group&Supervision or Detention Sentences&Total Sentences\\
    \midrule
    Light & 0 & 27\\
    Control & 4 & 30\\
    Harsh & 6 & 32\\
    
  \bottomrule
\end{tabular}
\end{table}

After completing the experiment, we conducted one-on-one interviews with all participants who had chosen to assign the punishment of supervision or detention. In the interviews, all ten participants stated that when conducting the experiment, they made judicial judgements with the punishment of supervision or detention influenced by cases retrieved. There are two thoughts among participant groups regarding this influence:

\begin{itemize}
    \item The first is that participants believe that there are some similarities between the cases retrieved and the task case's circumstances, according to which the sentence with similar circumstances should be assigned to the punishment of supervision or detention based on other circumstances that could reduce the punishment in the task case.
    \item The second is that participants believe that compared to the cases retrieved, the task case's circumstances are significantly less severe and have a smaller social influence. However, similar circumstances in the cases retrieved are sentenced to less than one year in prison, so the punishment of supervision or detention should be assigned to the task case.
\end{itemize}

Regardless of which approach, it shows that participants received influence from cases retrieved when making judicial judgements. From the perspective of criminal law sentences, supervision and detention are the lightest sentences for intentional injury crimes. If participants claimed that they did not receive influence from cases retrieved and only relied on the circumstances of the task case to make such a sentence, then it may be true that they did not consider cases retrieved deeply enough. However, if participants claimed that they considered cases retrieved before making non-imprisonable judicial judgements, then it can be argued that they made a leap from "could be reduced" to "should be reduced" due to the mitigating circumstances being statutory "could be reduced" rather than "should be reduced" and the original sentence being seven months in jail as a suspended sentence. Therefore, in this scenario, it can be believed that a reference case retrieval system influenced participants' decision-making process in judicial judgements.

\subsection{Questionnaire Results}

\begin{figure}
    \centering
    \includegraphics[width=\linewidth]{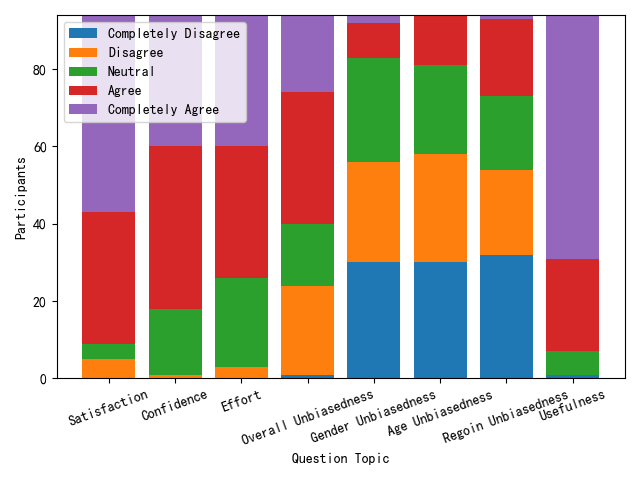}
    \caption{Questionnaire results}
    \label{fig:questionnaire result}
\end{figure}

The results of the 5-level user questionnaire are shown in \textbf{Figure~\ref{fig:questionnaire result}}. Most participants are generally satisfied with the experiment, had high confidence in their judicial judgement abilities, and found case studies to be helpful during the experimental process. Furthermore, since this experiment did not include any gender, age, or geographic bias in the case studies, most participants disagreed with the notion that such biases existed in the case studies, which is consistent with expectations.

Since most participants claim that the cases retrieved have assist their sentencing process, which supports the result of our user study, that is, shows that the sentences users make are influenced by our shown cases rather than decision-making by their knowledge and cognition only regardless of cases retrieved.
\subsection{Data Access}
The data used throughout the research and the user study results are available at: \href{https://github.com/Benson0704/LegalJudgeUserStudy}{https://github.com/Benson0704/LegalJudgeUserStudy}
\section{CONCLUSIONS}
Through statistical analysis of the experiment data, this study demonstrated that there is an influence of the case retrieval system on users. Specifically, the experiment showed that if participants are given reference to harsher sentences, they tended to make harsher judicial judgements, which is significant. However, if participants are given reference to lighter sentences, their tendency is not significant. Furthermore, through analyzing case-specific data from the experiment, it is found that the distribution of cases may affect users' decision-making process. The direction and pattern of influence need to be analyzed specifically and may contradict intuition.

This study still has limitations, such as that sentencing itself is a highly technical, practical, and subjective activity. Most law students lack practical experience and have not undergone standardized training in sentencing. In addition, sentencing is also affected by a series of uncontrollable factors such as the economic situation of the jurisdiction where the judiciary is located, criminal policies, and overall sentencing balance within a school. These factors are far beyond the capabilities of current experiment participants, which can adversely affect the validity and objectivity of the experiment.
\begin{acks}
    This work is supported by Quan Cheng Laboratory (Grant No. QCLZD202301)
    and the Natural Science Foundation of China (Grant No. 62002194).
\end{acks}


\bibliographystyle{ACM-Reference-Format}
\bibliography{sample-base}


\end{document}